\documentclass[twocolumn,showpacs,amsmath,amssymb]{revtex4}

\usepackage{graphicx}% Include figure files
\usepackage{graphics}% Include figure files
\usepackage{dcolumn}% Align table columns on decimal point
\usepackage{bm}% bold math
\def\rr{{{\bf r}}}
\def\rrp{{{\bf r}^\prime}}
\begin{document}

%\preprint{APS/123-QED}

\title{Self-consistent iterative solution of the exchange-only OEP equations for simple metal clusters in jellium model}

\author{M. Payami}
\affiliation{Center for Theoretical Physics and Mathematics,
Atomic Energy Organization of Iran, P. O. Box 11365-8486,
Tehran-Iran }
\author{Tahereh Mahmoodi}
\affiliation{Physics Department, Islamic Azad University, Branch
of Sciences and Research, Tehran}

\altaffiliation[Also at ]{Physics Department, Faculty of Sciences,
Islamic Azad University, Branch of Mash'had}

\date{\today}% It is always \today, today,

\begin{abstract}
In this work, employing the exchange-only orbital-dependent
functional, we have obtained the optimized effective potential
using the simple iterative method proposed by K\"ummel and Perdew
[S. K\"ummel and J. P. Perdew, Phys. Rev. Lett. {\bf 90}, 43004-1
(2003)]. Using this method, we have solved the self-consistent
Kohn-Sham equations for closed-shell simple metal clusters of Al,
Li, Na, K, and Cs in the context of jellium model. The results are
in good agreement with those obtained by the different method of
Engel and Vosko [E. Engel and S. H. Vosko, Phys. Rev. B {\bf 50},
10498 (1994)].
\end{abstract}

\pacs{71.15.-m, 71.15.Mb, 71.15.Nc, 71.20.Dg, 71.24.+q, 71.70.Gm}
\maketitle

\section{Introduction}\label{sec1}
In spite of the success of the local density approximation
(LDA)\cite{KS65} and the generalized gradient approximations
(GGA)\cite{Perdew85,Perdew96} for the exchange-correlation (XC)
part of the total energy in the density functional theory
(DFT)\cite{HK64}, it is observed that in some cases these
approximations lead to qualitatively incorrect results.
 On the other hand, appropriate
self-interaction corrected versions of these approximations are
observed\cite{Burke99} to lead to correct behaviors. These
observations motivate one to use functionals in which the
self-interaction contribution is removed exactly. One of the
functionals which satisfies this constraint is the exact exchange
energy functional. Using the exact exchange functional leads to
the correct asymptotic behavior of the Kohn-Sham (KS) potential as
well as to correct results for the high-density limit in which the
exchange energy is dominated. Given an orbital-dependent exchange
functional, one should solve the optimized effective potential
(OEP) integral equation\cite{Sharp,Talman,Sahni} to obtain the
local exchange potential which is used in the KS equations.
Application of this integral equation to three dimensional systems
\cite{Stadele99,Gorling99,Ivanov99} needs considerable
technicalities and has some limitations. Recently, K\"ummel and
Perdew \cite{KummelPRL03,KummelPRB03} proposed an iterative method
which allows one to solve the OEP integral equation accurately and
efficiently.

In this work, using the exact-exchange OEP method, we have
obtained the ground state properties of simple neutral
$N$-electron metal clusters of Al, Li, Na, K, and Cs with
closed-shell configurations corresponding to $N$= 2, 8, 18, 20,
34, and 40 (for Al, only $N=18$ corresponds to real Al cluster
with 6 atoms). However, it is a well-known fact that the
properties of alkali metals are dominantly determined by the
delocalized valence electrons. In these metals, the Fermi
wavelengths of the valence electrons are much larger than the
metal lattice constants and the pseudopotentials of the ions do
not significantly affect the electronic structure. This fact
allows one to replace the discrete ionic structure by a
homogeneous positive charge background which is called jellium
model (JM). For closed-shell clusters, the spherical geometry is
an appropriate assumption \cite{Payami99,Payami01,Payami04} and
therefore, we apply the JM to metal clusters by replacing the ions
of an $N$-atom cluster with a sphere of uniform positive charge
density and radius $R=(zN)^{1/3}r_s$, where $z$ is the valence of
the atom and $r_s$ is the bulk value of the Wigner-Seitz (WS)
radius for valence electrons. For Al, Li, Na, K, and Cs we take
$r_s$=2.07, 3.28, 3.93, 4.96, and 5.63, respectively.

The organization of this paper is as follows. In section
\ref{sec2} we explain the calculational schemes. Section
\ref{sec3} is devoted to the results of our calculations and
finally, we conclude this work in section \ref{sec4}.
%%%%%%%%%%%%%%%%%%%%%%%%%%%%%%%%%%%%%%%%%%%%%%%%%%%%%%%%%%%%%%%%%
\section{Calculational schemes}\label{sec2}
In the JM, the total energy of a cluster with exact exchange is
given by
\begin{eqnarray}\label{eq1}
\nonumber
E_{x-JM}[n_\uparrow,n_\downarrow,r_s]=T_s[n_\uparrow,n_\downarrow]+
   E_x[n_\uparrow,n_\downarrow]\\
   +\frac{1}{2}\int d\rr\;\phi([n,n_+];\rr)\,[n(\rr)-n_+(\rr)],
\end{eqnarray}
in which
\begin{equation}\label{eq2}
  E_x=\sum_{\sigma=\uparrow,\downarrow}\sum_{i,j=1}^{N_\sigma}
  \int d\rr \, d\rrp
  \frac{\phi_{i\sigma}^*(\rr)\phi_{j\sigma}^*(\rrp)
  \phi_{j\sigma}(\rr)\phi_{i\sigma}(\rrp)}{\mid\rr-\rrp\mid},
\end{equation}
and
\begin{equation}\label{eq3}
  \phi([n,n_+];\rr)=2\int
  d\rrp\frac{[n(\rrp)-n_+(\rrp)]}{\mid\rr-\rrp\mid}.
\end{equation}

Here, the background charge density is given by
\begin{equation}\label{eq4}
  n_+(\rr)=n\theta(R-r);\;\;\;\;\;\;n=\frac{3}{4\pi r_s^3}.
\end{equation}
and $n(\rr)$ is calculated from

\begin{equation}\label{eq5}
  n(\rr)=\sum_{\sigma=\uparrow,\downarrow}\sum_{i=1}^{N_\sigma}
  \mid\phi_{i\sigma}(\rr)\mid^2,
\end{equation}
where $\phi_{i\sigma}(\rr)$ are the KS orbitals obtained from the
self-consistent solutions of the set of equations
\begin{equation}\label{eq6}
(\hat{h}_{KS\sigma}-\varepsilon_{i\sigma})\phi_{i\sigma}(\rr)=0.
\end{equation}
In Eq.(\ref{eq6}),
\begin{equation}\label{eq7}
 \hat{h}_{KS\sigma}=-\nabla^2+v_{eff\sigma}(\rr),
\end{equation}
\begin{equation}\label{eq8}
 v_{eff\sigma}(\rr)=v(\rr)+v_H(\rr)+v_{x\sigma}(\rr)
\end{equation}
\begin{equation}\label{eq9}
  v_H(\rr)=2\int d\rr\,\frac{n(\rrp)}{\mid\rr-\rrp\mid}.
\end{equation}
 All equations throughout this paper are expressed in Rydberg
atomic units.

To solve the KS equations, one should first calculate the local
exchange potential from the exchange energy functional. This is
done via the solution of the OEP integral equation. Recently,
K\"ummel and Perdew\cite{KummelPRL03,KummelPRB03} in a simple and
elegant way have proved that the OEP integral equation is
equivalent to the equation
\begin{equation}\label{eq10-1}
  \sum_{i=1}^{N_\sigma}\psi_{i\sigma}^*(\rr)\phi_{i\sigma}(\rr)+c.c.=0,
\end{equation}
in which $\phi_{i\sigma}$ are the self-consistent KS orbitals and
$\psi_{i\sigma}$ are orbital shifts which are obtained from the
solution of the following inhomogeneous KS equations
\begin{equation}\label{eq10}
(\hat{h}_{KS\sigma}-\varepsilon_{i\sigma})\psi_{i\sigma}^*(\rr)=Q_{i\sigma}(\rr),
\end{equation}
with
\begin{equation}\label{eq10-2}
Q_{i\sigma}(\rr)=-[v_{x\sigma}(\rr)-u_{xi\sigma}(\rr)-(\bar{v}_{xi\sigma}
  -\bar{u}_{xi\sigma})]\phi_{i\sigma}^*(\rr).
\end{equation}

$\varepsilon_{i\sigma}$ are the KS eigenvalues which satisfy Eq.
(\ref{eq6}), and in the right hand side of Eq. (\ref{eq10-2}),
$v_{x\sigma}(\rr)$ are the optimized effective potential and
\begin{equation}\label{eq11}
  u_{xi\sigma}(\rr)=-\frac{2}{\phi_{i\sigma}^*(\rr)}\sum_{j=1}^{N_\sigma}
  \phi_{j\sigma}^*(\rr)
  \int
  d\rrp\frac{\phi_{i\sigma}^*(\rrp)\phi_{j\sigma}(\rrp)}{\mid\rr-\rrp\mid},
\end{equation}
\begin{equation}\label{eq12}
  \bar{v}_{xi\sigma}=\int
  d\rr\phi_{i\sigma}^*(\rr)v_{x\sigma}(\rr)\phi_{i\sigma}(\rr),
\end{equation}
\begin{equation}\label{eq13}
 \bar{u}_{xi\sigma}=\int
  d\rr\phi_{i\sigma}^*(\rr)u_{xi\sigma}(\rr)\phi_{i\sigma}(\rr).
\end{equation}

 At the starting point to solve the self-consistent OEP equations (\ref{eq10})-(\ref{eq13}),
the self-consistent KLI \cite{KLI}  orbitals and eigenvalues are
used as input.
  Then we solve Eq. (\ref{eq10}) to
obtain the orbital shifts $\psi_{i\sigma}$. In the next step, we
calculate the quantity
\begin{equation}\label{eq14}
  S_\sigma(\rr)=\sum_{i=1}^{N_\sigma}\psi_{i\sigma}^*(\rr)\phi_{i\sigma}(\rr)
  +c.c.,
\end{equation}
the deviation of which from zero is a measure for the deviation
from the self-consistency of the OEP-KS orbitals. This quantity is
used to construct a better exchange potential from
\begin{equation}\label{eq15}
  v_{x\sigma}^{new}(\rr)=v_{x\sigma}^{old}(\rr)+cS_\sigma(\rr).
\end{equation}

With this $v_{x\sigma}^{new}(\rr)$ and keeping the KS eigenvalues
and orbitals fixed, we repeat the solution of the Eq.
(\ref{eq10}). Repeating the "cycle" (\ref{eq10}), (\ref{eq14}),
(\ref{eq15}) for several times, the maximum value of
$S_\sigma(\rr)$ will decrease to a desired small value (in our
case down to $10^{-8}$ a. u.). After completing cycles, the
$v_{x\sigma}^{new}$ in conjunction with the KS orbitals are used
to construct new effective potential to "iterate" the KS equations
(\ref{eq6}). The value of $c$ in Eq. (\ref{eq15}) is taken to be
30 as suggested in Ref.\cite{KummelPRB03}. We have used 10 cycles
between two successive iterations. These procedures are repeated
until the self-consistent OEP potentials are obtained.
%%%%%%%%%%%%%%%%%%%%%%%%%%%%%%%%%%%%%%%%%%%%%%%%%%%%%%%%%%%%%%%
\section{Results and discussion}\label{sec3}
Taking spherical geometry for the jellium background, and solution
of self-consistent KS equations, we have obtained the ground state
properties of closed-shell 2, 8, 18, 20, 34, and 40-electron
neutral clusters of Al, Li, Na, K, and Cs in the exact-exchange
jellium model and compared the results with those of KLI and LSDA.

To solve the KS and OEP equations for spherical geometry we take
\begin{equation}\label{eq16}
  \phi_{i\sigma}(\rr)=\frac{\chi_{i\sigma}(r)}{r}Y_{l_i,m_i}(\Omega)
\end{equation}
and
\begin{equation}\label{eq17}
  \psi_{i\sigma}(\rr)=\frac{\xi_{i\sigma}(r)}{r}Y_{l_i,m_i}(\Omega).
\end{equation}

Substitution of Eq. (\ref{eq16}) and Eq. (\ref{eq17}) into Eq.
(\ref{eq10}) the inhomogeneous KS equation reduces to

\begin{equation}\label{eq18}
  \left[\frac{d^2}{dr^2}+\varepsilon_{i\sigma}-v_{eff\sigma}(r)-\frac{l_i(l_i+1)}{r^2}\right]\xi_{i\sigma}(r)
  =q_{i\sigma}(r),
\end{equation}
in which

\begin{equation}\label{eq19}
  q_{i\sigma}(r)=q_{i\sigma}^{(1)}(r)+q_{i\sigma}^{(2)}(r),
\end{equation}
with
\begin{equation}\label{eq20}
  q_{i\sigma}^{(1)}(r)=\left[v_{xc\sigma}(r)-\bar
  v_{xci\sigma}+\bar u_{xci\sigma}\right]\chi_{i\sigma}(r),
\end{equation}
and
\begin{eqnarray}\label{eq21}
q_{i\sigma}^{(2)}(r)=2\sum_{j=1}^{N_\sigma}\sum_{l=|l_i-l_j|}^{l_i+l_j}\frac{4\pi}{2l+1}\chi_{j\sigma}(r)B_\sigma(i,j,l;r)
\nonumber\\
\times\overline{\left[I(l_jm_j,l_im_i,lm_j-m_i)\right]^2}.
\end{eqnarray}

The quantities $B$ and $I$ in Eq. (\ref{eq21}) are defined as
\begin{eqnarray}\label{eq22}
  B_\sigma(i,j,l;r)=\int_{r^\prime=0}^rdr^\prime\chi_{i\sigma}(r^\prime)\chi_{j\sigma}(r^\prime)\frac{{r^\prime}^l}{r^{l+1}}
  \nonumber\\
  +\int_{r^\prime=r}^\infty dr^\prime\chi_{i\sigma}(r^\prime)\chi_{j\sigma}(r^\prime)\frac{r^l}{{r^\prime}^{l+1}}
\end{eqnarray}
\begin{equation}\label{eq23}
  I(l_jm_j,l_im_i,lm)=\int d\Omega\;
  Y_{l_jm_j}^*(\Omega)Y_{l_im_i}(\Omega)Y_{lm}(\Omega),
\end{equation}
and the bar over $I^2$ implies average over $m_i$ and $m_j$.
Also, the expression for $\bar u_{xi\sigma}$ reduces to
\begin{eqnarray}\label{eq21-1}
\bar
u_{xi\sigma}=-2\sum_{j=1}^{N_\sigma}\sum_{l=|l_i-l_j|}^{l_i+l_j}\frac{4\pi}{2l+1}
\overline{\left[I(l_jm_j,l_im_i,lm_j-m_i)\right]^2}\nonumber \\
\times\int_0^\infty dr\;\chi_{i\sigma}(r)\chi_{j\sigma}(r)
B_\sigma(i,j,l;r).\;\;\;
\end{eqnarray}

\begin{figure}[htb]
\includegraphics[width=8cm]{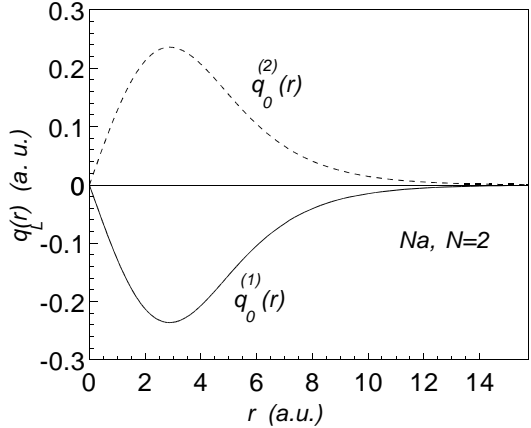}
\caption{\label{fig1} Source terms $q_{l=0,\sigma}^{(1)}$ and
$q_{l=0,\sigma}^{(2)}$ in atomic units for Na$_2$. As is obvious,
the two terms are equal and opposite in sign, so that the orbital
shift for $N=2$ vanishes and the KLI and OEP results coincide.}
\end{figure}

 In Fig.\ref{fig1}, the source term components $q_{l=0,\sigma}^{(1)}$ and
$q_{l=0,\sigma}^{(2)}$ are plotted as functions of radial
coordinate. As is seen, they are equal and opposite in sign so
that they lead to zero orbital shift, i.e.,
$\xi_{l=0,\sigma}(r)=0$. This result in turn leads to the
coincidence of the KLI and OEP results.

\begin{figure}[htb]
\includegraphics[width=17cm]{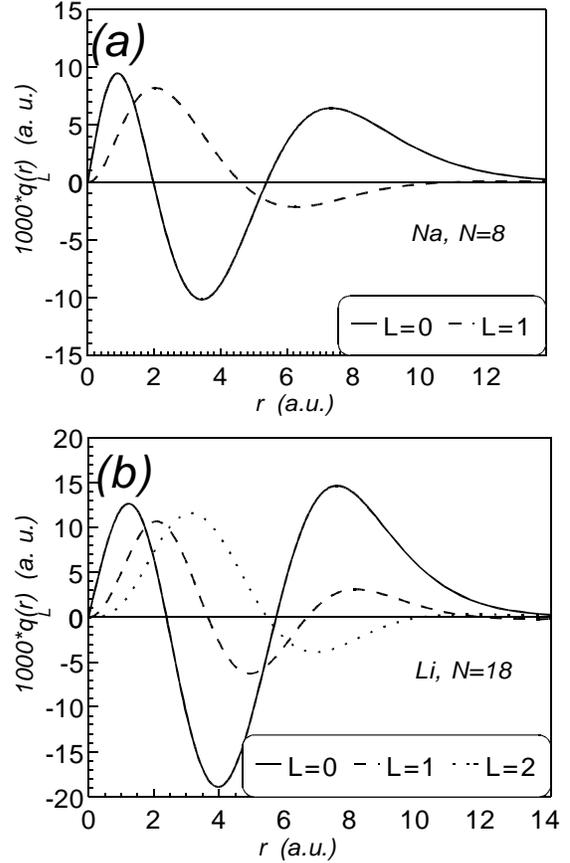}
\caption{\label{fig2} (a)-Source terms $q_{l\sigma}$ in atomic
units for Na$_8$, and (b)- for Li$_{18}$. For Na$_8$, only $l=0$
and $l=1$ orbitals are occupied for each spin component whereas,
for Li$_{18}$, the orbitals with $l=0,1,2$ are occupied.}
\end{figure}

\begin{figure}[htb]
\includegraphics[width=17cm]{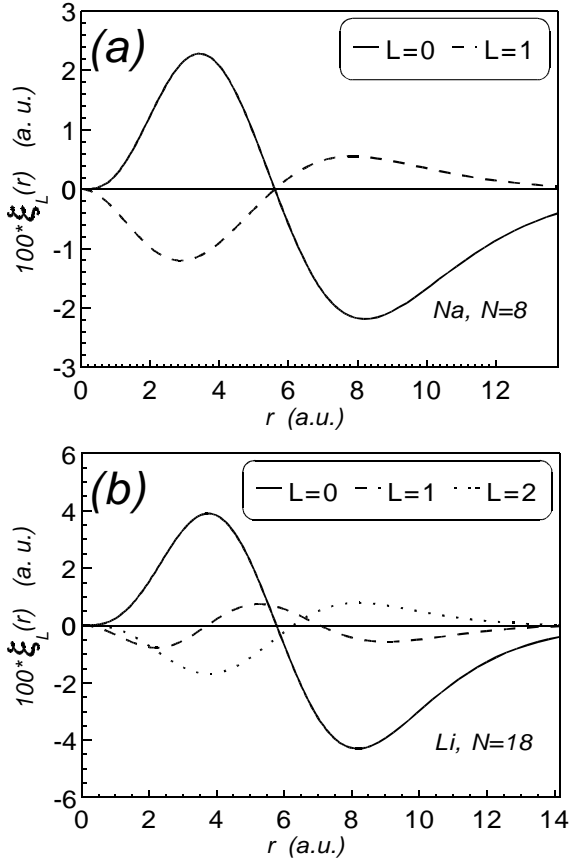}
\caption{\label{fig3} Orbital shifts $\xi_{l\sigma}$ in atomic
units (a)- for Na$_8$ and (b)- for Li$_{18}$.}
\end{figure}

In Figs. \ref{fig2}(a) and \ref{fig2}(b) the self-consistent
source terms $q_{l\sigma}(r)$ of Eq.(\ref{eq20}) are plotted as
functions of radial coordinate for Na$_8$ and Li$_{18}$,
respectively. The corresponding orbital shifts $\xi_{l\sigma}$ are
shown in Figs. \ref{fig3}(a) and \ref{fig3}(b). It should be noted
that $q_{i\sigma}(r)$ and $\xi_{i\sigma}(r)$ must behave such that
\begin{equation}\label{eq25}
  \int d\rr\;Q_{i\sigma}(\rr)\phi_{i\sigma}(\rr)=0
\end{equation}
and
\begin{equation}\label{eq26}
  \int d\rr\;\psi^*_{i\sigma}(\rr)\phi_{i\sigma}(\rr)=0
\end{equation}
are satisfied.

In order to solve the self-consistent OEP equations, we use the
KLI self-consistent results as input. For the KLI calculations, we
use [Eq.(23) of Ref.\cite{KummelPRB03} with
$\psi_{i\sigma}(\rr)=0$]:
\begin{eqnarray}\label{eq24}
  v_{x\sigma}^{\rm KLI}(\rr)=\frac{1}{2n_\sigma(\rr)}\sum_{i=1}^{N_\sigma}\left\{\phi_{i\sigma}(\rr)\phi_{i\sigma}^*(\rr)
  u_{xi\sigma}(\rr)\right.\nonumber \\
   \left.+ |\phi_{i\sigma}(\rr)|^2(\bar{v}_{xi\sigma}-\bar{u}_{xi\sigma})\right\}
   +c.c.
\end{eqnarray}

\begin{figure}[htb]
\includegraphics[width=17cm]{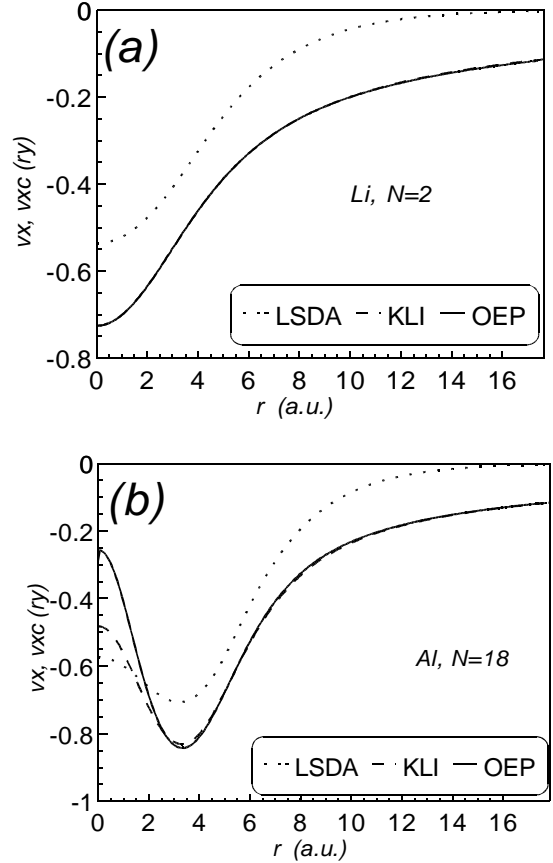}
\caption{\label{fig4} Exchange potentials in KLI and OEP and
exchange-correlation potentials in LSDA, in Rydbergs, for (a)-
Li$_2$ and, (b)- for Al$_{18}$. Here 18 refers to the electrons
which is equivalent to 6 Al atoms. In Li$_2$ the KLI and OEP
completely coincide whereas, in Al$_{18}$ the coincidence occurs
at asymptotic region. The LSDA has wrong exponential decay
whereas, KLI and OEP have correct $1/r$ decays.}
\end{figure}

The self-consistent exchange potentials of Li$_2$ and Al$_{18}$
are plotted in Figs. \ref{fig4}(a) and \ref{fig4}(b),
respectively. For comparison, the LSDA exchange-correlation
potentials are also included. One notes that in Li$_2$ case, the
KLI and OEP potentials are completely coincident whereas, in Al
case, the KLI and OEP coincide only in the asymptotic region. On
the other hand, the LSDA potential, because of wrong exponential
asymptotic behavior, decays faster than the KLI or OEP, which have
correct asymptotic behaviors of $1/r$. In the Al case, $N=18$
refers to the number of electrons which corresponds to the number
$n=6$ of Al atoms.

\begin{figure}[htb]
\includegraphics[width=17cm]{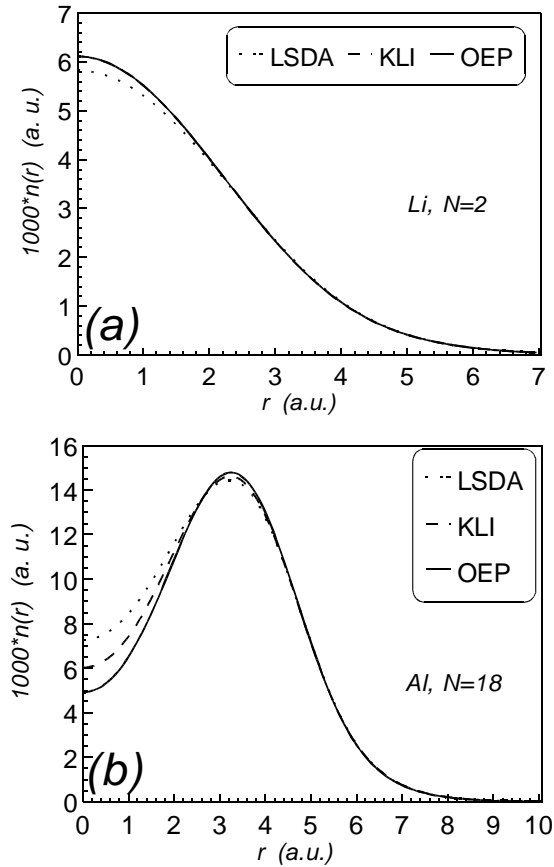}
\caption{\label{fig5}Densities in atomic units for Li$_2$ and,
(b)- for Al$_{18}$. Here, as in potentials, we have full
coincidence for Li$_2$ and asymptotic coincidence for Al$_{18}$.}
\end{figure}

In Figs. \ref{fig5}(a) and \ref{fig5}(b), we have shown the
self-consistent densities for Li$_2$ and Al$_{18}$, respectively.
As in the potential case, for Li$_2$ the KLI and OEP densities
completely coincide whereas, in Al$_{18}$ the coincidence is only
at the asymptotic region.

In Table \ref{table1} we have listed the self-consistent
calculated ground state properties of the closed-shell clusters of
Al, Li, Na, K, Cs for $N=$2, 8, 18, 20, 34, and 40. For comparison
of our OEP results with those obtained by Engel and Vosko
(EV)\cite{EV}, we have also included those results for Al, Na, and
Cs. The EV results are based on gradient expansion which, in
principle, is valid only for slow variations of density as in a
bulk solid. However, for finite systems such as clusters or
surfaces, the EV results may differ from the exact OEP results.
Comparison of our OEP total energies with those of EV for Na
clusters shows a difference of $0.002\%$ on average. On the other
hand, the EV exchange energies differ, on average, by $0.001\%$
and the average difference in $\varepsilon_H$ is $0.08\%$. From
the computational costs point of view, these quite small
differences makes the EV method advantageous for calculations
within above mentioned accuracies.

Now we compare the total energies and the exchange energies in the
KLI, OEP, and LSDA schemes. Comparison of the total energies shows
that the OEP energies, on the average, are $1.2\%$ less than those
of the KLI. We do not compare the total energies of OEP and LSD
because in LSD there exist a correlation contribution. On the
other hand, comparison of the exchange energies shows that on the
average, the exchange energies in the OEP is $0.33\%$ more
negative than that of the KLI whereas, it is $9\%$ more negative
than the LSD.

 An other feature in OEP which should be noted is the
contraction of the KS eigenvalue bands relative to those of KLI.
The results in Table \ref{table1} show that for all $N$, the
relation $\Delta^{\rm OEP}<\Delta^{KLI}$ holds. Here,
$\Delta=\varepsilon_H-\varepsilon_L$ is the difference between the
maximum occupied and minimum occupied KS eigenvalues. For $N$=2,
we have $\Delta=0$. The results show that the maximum relative
contraction,$|\Delta^{\rm OEP}-\Delta^{KLI}|/\Delta^{KLI}$, is
2.6$\%$ which corresponds to Cs$_{18}$.

\begingroup
\squeezetable
\begin{table*}
\caption{\label{table1}Absolute values of total and exchange
energies as well as highest occupied and lowest occupied Kohn-Sham
eigenvalues in Rydbergs. The LSDA total energies include the
correlation energies as well.}
\begin{ruledtabular}
\begin{tabular}{cccccccccccccccccc}
& & &\multicolumn{4}{c}{LSDA}&\multicolumn{4}{c}{KLI}&\multicolumn{4}{c}{OEP}&\multicolumn{3}{c}{EV\footnote{Data from Ref.\cite{EV}.}}\\
 Atom&$r_s$&$N$&$-E$&$-E_x$&$-\varepsilon_L$&$-\varepsilon_H$&$-E$&$-E_x$&$-\varepsilon_L$&$-\varepsilon_H$&
 $-E$&$-E_x$&$-\varepsilon_L$&$-\varepsilon_H$&$-E$&$-E_x$&$-\varepsilon_H$\\ \hline
Al\footnote{Here, $N$=18 corresponds to Al$_6$ cluster and other
$N$'s do not correspond to a real Al clusters.}
&2.07&2&0.0944&0.5936&0.3821&0.3821&0.0557&0.7016&0.5973&0.5973&0.0557&0.7016&0.5973&0.5973&0.0557&0.7016&0.5973 \\
    &&8&0.3087&2.7822&0.6957&0.3806&-0.0660&3.0178&0.8552&0.5418&-0.0653&3.0248&0.8507&0.5416&-0.0653&3.0248&0.5417 \\
   &&18&0.4519&6.6899&0.8606&0.3411&-0.6023&7.0693&0.9710&0.4618&-0.5998&7.0987&0.9608&0.4600&-0.5998&7.0987&0.4600 \\
   &&20&0.6444&7.4183&0.8556&0.3215&-0.5493&7.7898&0.9662&0.4333&-0.5480&7.8071&0.9618&0.4326&-0.5480&7.8071&0.4316 \\
   &&34&0.7603&13.1379&0.9522&0.3103&-1.4409&13.7043&1.0356&0.4066&-1.4354&13.7536&1.0298&0.4027&-1.4354&13.7535&0.4027 \\
   &&40&1.0806&15.3585&0.9497&0.3082&-1.6022&15.8635&1.0369&0.3996&-1.6000&15.8913&1.0307&0.3956&-1.6001&15.8913&0.3955 \\
&&&&&&&&&&&&&&&&& \\
Li
&3.28&2&0.2327&0.4324&0.2736&0.2736&0.1866&0.5074&0.4203&0.4203&0.1866&0.5074&0.4203&0.4203&-&-&- \\
    &&8&1.0141&1.9015&0.4074&0.2752&0.6708&2.0538&0.5097&0.3779&0.6714&2.0591&0.5076&0.3781&-&-&- \\
   &&18&2.3050&4.4733&0.4714&0.2598&1.3930&4.7233&0.5404&0.3338&1.3952&4.7474&0.5352&0.3328&-&-&- \\
   &&20&2.6056&4.9417&0.4681&0.2303&1.5677&5.1710&0.5316&0.2992&1.5689&5.1842&0.5295&0.3000&-&-&- \\
   &&34&4.4619&8.6619&0.5065&0.2494&2.5990&9.0347&0.5570&0.3061&2.6040&9.0778&0.5533&0.3037&-&-&- \\
   &&40&5.2635&10.1016&0.5014&0.2267&2.9843&10.3981&0.5491&0.2794&2.9865&10.4195&0.5464&0.2783&-&-&- \\
&&&&&&&&&&&&&&&&& \\
Na
&3.93&2&0.2462&0.3787&0.2381&0.2381&0.1988&0.4428&0.3627&0.3627&0.1988&0.4428&0.3627&0.3627&0.1988&0.4428&0.3626 \\
    &&8&1.0737&1.6290&0.3333&0.2402&0.7465&1.7551&0.4177&0.3249&0.7470&1.7598&0.4162&0.3251&0.7470&1.7598&0.3252 \\
   &&18&2.4664&3.8049&0.3777&0.2297&1.6128&4.0135&0.4338&0.2896&1.6148&4.0354&0.4298&0.2888&1.6148&4.0354&0.2888 \\
   &&20&2.7664&4.1991&0.3748&0.2018&1.7944&4.3852&0.4250&0.2577&1.7956&4.3974&0.4237&0.2588&1.7956&4.3974&0.2600 \\
   &&34&4.7746&7.3347&0.4022&0.2232&3.0446&7.6461&0.4424&0.2679&3.0493&7.6870&0.4392&0.2659&3.0494&7.6870&0.2662 \\
   &&40&5.6075&8.5495&0.3976&0.2002&3.4899&8.7840&0.4337&0.2412&3.4920&8.8038&0.4320&0.2410&3.4920&8.8036&0.2414 \\
&&&&&&&&&&&&&&&&& \\
K
&4.96&2&0.2448&0.3174&0.1981&0.1981&0.1970&0.3693&0.2979&0.2979&0.1970&0.3693&0.2979&0.2979&-&-&- \\
    &&8&1.0596&1.3306&0.2594&0.2006&0.7553&1.4280&0.3245&0.2658&0.7557&1.4319&0.3235&0.2660&-&-&- \\
   &&18&2.4442&3.0822&0.2874&0.1943&1.6667&3.2447&0.3294&0.2389&1.6685&3.2639&0.3266&0.2383&-&-&- \\
   &&20&2.7275&3.3986&0.2851&0.1700&1.8420&3.5380&0.3214&0.2120&1.8431&3.5490&0.3211&0.2134&-&-&- \\
   &&34&4.7230&5.9117&0.3030&0.1908&3.1617&6.1552&0.3320&0.2229&3.1662&6.1934&0.3295&0.2214&-&-&- \\
   &&40&5.5338&6.8879&0.2995&0.1701&3.6226&7.0565&0.3234&0.1988&3.6247&7.0744&0.3230&0.1994&-&-&- \\
&&&&&&&&&&&&&&&&& \\
Cs
&5.63&2&0.2382&0.2875&0.1789&0.1789&0.1907&0.3335&0.2669&0.2669&0.1907&0.3335&0.2669&0.2669&0.1907&0.3335&0.2669 \\
    &&8&1.0252&1.1904&0.2271&0.1816&0.7341&1.2742&0.2833&0.2376&0.7345&1.2778&0.2826&0.2378&0.7345&1.2777&0.2378 \\
   &&18&2.3652&2.7459&0.2490&0.1768&1.6290&2.8866&0.2846&0.2144&1.6307&2.9044&0.2823&0.2139&1.6307&2.9043&0.2132 \\
   &&20&2.6351&3.0268&0.2471&0.1548&1.7969&3.1446&0.2772&0.1904&1.7980&3.1553&0.2773&0.1920&1.7980&3.1553&0.1925 \\
   &&34&4.5646&5.2538&0.2613&0.1743&3.0932&5.4652&0.2851&0.2007&3.0974&5.5020&0.2830&0.1994&3.0974&5.5020&0.1974 \\
   &&40&5.3452&6.1206&0.2584&0.1554&3.5445&6.2591&0.2770&0.1787&3.5462&6.2788&0.2766&0.1791&3.5465&6.2763&0.1795 \\
\end{tabular}
\end{ruledtabular}
\end{table*}
\endgroup
\section{Summary and Conclusion}\label{sec4}
In this work, we have considered the exchange-only jellium model
in which we have used the exact orbital-dependent exchange
functional. This model is applied for the closed-shell simple
metal clusters of Al, Li, Na, K, and Cs. For the local exchange
potential in the KS equation, we have solved the OEP integral
equation by the iterative method proposed recently by K\"ummel and
Perdew \cite{KummelPRB03}. By solving the self-consistent KS
equations, we have obtained the ground state energies of the
closed-shell clusters ($N=2,8,18,20,34,40$) for the three schemes
of LSD, KLI, and OEP. The KLI and OEP results are the same for
neutral two-electron clusters. However, for $N\neq 2$, the
densities and potentials in the KLI and OEP coincide for large $r$
values. The OEP exchange and effective potentials shows correct
behavior of $1/r$ compared to the incorrect exponential behavior
in the LSD. The total energies in the OEP are more negative than
the KLI by $1.2\%$ on the average. On the other hand, the exchange
energies in the OEP is about $0.33\%$ more negative than that in
the KLI whereas, it is about $9\%$ more negative than that in the
LSDA. The widths of the occupied bands,
$\varepsilon_H-\varepsilon_L$ in the OEP are contracted relative
to those in the KLI by at most $2.6\%$ which corresponds to
Cs$_{18}$. In spite of the validity of the gradient expansion
method for slow variations in density, comparison of our OEP
results with those of EV shows an excellent agreement.

\acknowledgements{ M. P. would like to appreciate the useful
comments of Prof. John P. Perdew. Also, he would like to thank
Prof. Eberhard Engel for providing the unpublished results on Al
and Cs.}
%\newpage %Just because of unusual number of tables stacked at end

\end{document}